\documentclass[prl,reprint,superscriptaddress,preprintnumbers]{revtex4-1}
\pdfoutput=1
\usepackage{hyperref}
\usepackage[T1]{fontenc}

\usepackage{mathrsfs}

\usepackage{amsmath,amssymb,amsfonts,amsxtra,mathrsfs,graphics,graphicx,amsthm,epsfig,bm,longtable,float,color,tikz,mathtools,xfrac,footnote}
\restylefloat{table}
\usepackage{dsfont}
\usepackage{lipsum}
\usepackage{textgreek}
\usetikzlibrary{decorations.pathmorphing}
\usetikzlibrary{decorations.markings}
\usetikzlibrary{quotes,arrows.meta}
\usetikzlibrary{arrows,decorations.markings,calc,fadings,decorations.pathreplacing,patterns,decorations.pathmorphing,positioning}
\usepackage{tikz-cd}

\newcommand{\ii}{\mathrm{i}}

\newcommand{\be}{\begin{equation}} \newcommand{\ee}{\end{equation}}
\newcommand{\bea}{\begin{equation} \begin{aligned}} \newcommand{\eea}{\end{aligned} \end{equation}}

\def\U{\mathrm{U}}

\def\SU{\mathrm{SU}}

\newcommand{\rd}{\mathrm{d}}

\newcommand{\Vol}{\mathrm{Vol}}

\newcommand{\wh}{\widehat}

\newcommand{\pd}{\partial}

\newcommand{\cI}{\mathcal{I}}

\newcommand{\cN}{\mathcal{N}}

\newcommand{\cP}{\mathcal{P}}

\newcommand{\cV}{\mathcal{V}}

\newcommand{\bC}{\mathbb{C}}

\newcommand{\bR}{\mathbb{R}}
\newcommand{\bV}{\mathbb{V}}
\newcommand{\bZ}{\mathbb{Z}}

\newcommand{\fg}{\mathfrak{g}}

\newcommand{\fn}{\mathfrak{n}}

\newcommand{\fp}{\mathfrak{p}}
\newcommand{\fq}{\mathfrak{q}}

\begin{document}

\title{$\mathcal{I}$-extremization with baryonic charges}

\date{\today}

\author{Seyed Morteza Hosseini}
\affiliation{Centre for Theoretical Physics, Department of Physics and Astronomy, Queen Mary University of London, London E1 4NS, UK}
\email{morteza.hosseini@qmul.ac.uk}

\author{Alberto Zaffaroni}
\affiliation{Dipartimento di Fisica, Universit\`a di Milano-Bicocca, I-20126 Milano, Italy}
\affiliation{INFN, sezione di Milano-Bicocca, I-20126 Milano, Italy}

\email{alberto.zaffaroni@mib.infn.it}

\begin{abstract}

We propose an entropy function for AdS$_4$ BPS black holes in M-theory with general magnetic charges, resolving in particular a long-standing puzzle about baryonic charges. The entropy function is constructed from a gravitational block defined solely in terms of topological data of the internal manifold. We show that the entropy of twisted black holes can always be reformulated as an $\mathcal{I}$-extremization problem---even in cases where existing large-$N$ field theory computations fail to provide an answer. Furthermore, we correctly reproduce the entropy for a class of known black holes with purely baryonic magnetic charges. Our results offer both a conjecture for the general gravitational block for AdS$_4$ black holes in M-theory and a prediction for the large-$N$ limit of several partition functions whose saddle points have yet to be found. 

\end{abstract}



\maketitle

\emph{\textbf{Introduction.---}}
The field theory derivation of the entropy of supersymmetric black holes in anti de Sitter (AdS) spacetime is a remarkable success of holography. For a class of magnetically charged AdS$_4$ black holes, this is achieved by extremizing the large-$N$ limit of the topologically twisted index of the dual field theory \cite{Benini:2015noa}. This principle, known as $\cI$-extremization, was first applied to black holes in AdS$_4 \times S^7$ in \cite{Benini:2015eyy}. More precisely, for magnetically charged black holes with horizon the Riemann surface $\Sigma_{\fg}$, the quantity to be extremized is the logarithm of the partition function, $\cI = \log Z_{\Sigma_{\fg} \times S^1}$, of the dual three-dimensional $\cN = 2$ superconformal field theory (SCFT) on $\Sigma_{\fg} \times S^1$, with a topological twist along $\Sigma_{\fg}$. The quantity $\cI( \Delta_a, \fn_a)$ depends on a set of chemical potentials $\Delta_a$ and a set of magnetic fluxes $\fn_a$ along $\Sigma_{\fg}$ for the global symmetries of the theory. Interestingly, for a large class of $\cN = 2$ three-dimensional quiver gauge theories with a holographic dual in M-theory or massive IIA, the index at large $N$ takes the form \cite{Hosseini:2016tor}
\bea
 \label{II}
 \cI (\Delta_a, \fn_a) = - \frac 12 \sum_a \fn_a \frac{\pd F_{S^3}(\Delta_a)}{\pd \Delta_a} \, ,
\eea
where $F_{S^3}(\Delta_a)$ is the three-sphere partition function. This formula has a generalization to more general dyonic, accelerating, and rotating black holes by considering the appropriate fugacities and supersymmetric partition functions. There is indeed strong evidence that all the corresponding supersymmetric indices can be expressed at large $N$ in terms of the $S^3$ partition function of the dual theory. This follows from the holomorphic block factorization of three-dimensional partition functions \cite{Pasquetti:2011fj,Beem:2012mb} and has been confirmed at large $N$ in many examples \cite{Choi:2019dfu,Hosseini:2022vho,Colombo:2024mts}.

There is a gravitational side of the story. It was conjectured in \cite{Hosseini:2019iad} that the entropies of all black holes associated with a given dual theory can be found by extremizing \emph{entropy functions} obtained by gluing gravitational blocks. For three-dimensional theories, the gravitational block can be identified with the $S^3$ partition function, when written in the right variables. For a class of solutions, the entropy functions can be explicitly derived using the method of \cite{Couzens:2018wnk,Gauntlett:2018dpc} and defined as supersymmetric actions $S(b, \lambda_a)$, which depend on the supersymmetric Killing vector $b$ (Reeb vector) and a set of K\"ahler parameters $\lambda_a$ of the internal manifolds, subject to a set of constraints. The extremization problems in field theory and gravity are expected to be equivalent, though with a nontrivial mapping of parameters. This equivalence can be proven in several cases \cite{Hosseini:2019use,Hosseini:2019ddy,Gauntlett:2019roi,Boido:2022mbe}.

And here comes a puzzle. The global symmetries of theories dual to AdS$_4 \times \text{SE}_7$ solutions in M-theory, where $\text{SE}_7$ is a seven-dimensional Sasaki-Einstein manifold, can be divided into \emph{mesonic}, associated with the isometries of the internal manifold, and \emph{baryonic}, associated with gauge fields obtained by reducing antisymmetric forms on nontrivial cycles. It turns out that large-$N$ saddle points for the $S^3$ free energy and  other supersymmetric indices, beside existing only for a subclass of dual SCFTs, do not depend on the baryonic fugacities and fluxes \cite{Jafferis:2011zi,Hosseini:2016tor,Hosseini:2016ume,Choi:2019dfu}. The equivalence of $\cI$-extremization and the gravitational extremization of \cite{Couzens:2018wnk,Gauntlett:2018dpc} has been proven only for a choice of purely mesonic fluxes, dubbed the mesonic twist \cite{Hosseini:2019ddy,Gauntlett:2019roi}. This contrasts with the case of four-dimensional SCFTs associated with five-dimensional Sasaki-Einstein manifolds, where the analogue of $F_{S^3}(\Delta_a)$ is the trial central charge $a (\Delta_a)$, which depends on all the R-charges, and the equivalence of $c$-extremization \cite{Benini:2012cz,Benini:2013cda} and its gravitational dual can be proven for all choices of fluxes \cite{Hosseini:2019use}. Also in three dimensions, we expect that the entropy of the most general black hole depends on \emph{all} fluxes, including the baryonic ones. This is confirmed by the existence of explicit black hole solutions with baryonic charges \cite{Halmagyi:2013sla,Azzurli:2017kxo,Hong:2019wyi,Kim:2020qec}, more generally, by the method of \cite{Couzens:2018wnk,Gauntlett:2018dpc}, which depends on purely topological data and can compute the entropy even when the solution is not known.

In this letter, we show how to define a generalization of the large-$N$ free energy $F_{S^3}(\Delta_a)$ in terms of the master volume \cite{Gauntlett:2018dpc} of the internal manifold $\text{SE}_7$. We also argue that, quite remarkably, \emph{all} the supersymmetric conditions \cite{Couzens:2018wnk,Gauntlett:2018dpc} for the most general static magnetically charged black holes in AdS$_4 \times \text{SE}_7$ are equivalent to the extremization of \eqref{II}. We thus reformulate the extremization problem in the presence of baryonic charges in the  language of $\cI$-extremization, completing the analysis of \cite{Hosseini:2019ddy,Gauntlett:2019roi} and providing a prediction for the large-$N$ limit of several partition functions whose saddle points have yet to be found. Due to the technical complexity of the equations, our analysis is based on examples, but we believe the result to be valid in general (see \cite{Hosseini:2025mgf}). We propose that the generalized $F_{S^3}(\Delta_a)$ is the \emph{gravitational block} that should be used to construct entropy functions for the most general AdS$_4$ black holes, including Kerr-Newman and accelerating ones \cite{Ferrero:2020twa}. Without further ado, we introduce the characters of the play and formulate our main result.

\emph{\textbf{The extremization problem.---}}
We first consider AdS$_4 \times \text{SE}_7$ solutions corresponding to a set of $N$ M2-branes probing a seven-dimensional Sasaki-Einstein manifold $\text{SE}_7$. We also assume that $\text{SE}_7$ is toric and that the corresponding cone $C ( \text{SE}_7)$ is a Calabi-Yau (CY) four-fold defined by a fan with vectors $v^a$, $a = 1, \ldots, d$, whose first component is equal to one. The CY can be realized as the symplectic reduction of $\bC^d$ with respect to the subgroup $( \bC^*)^{d-4}$, whose generators $B^{(r)}$ satisfy $\sum_{a = 1}^{d} B_a^{(r)} v^a = 0$, for $r = 1, \dots, d - 4$. From this realization, we see that the global symmetry of the theory is the isometry group $\U(1)^d$ of $\bC^d$, which can be split into a $\U(1)^4$ mesonic subgroup corresponding to the isometries of $\text{SE}_7$ and $d - 4$ baryonic $\U(1)$'s. The $\U(1)$ symmetries of the theory can then be naturally labeled by the index $a$. We formulate our extremization problem by deforming the solution, allowing the Reeb vector $b = \sum_{i=1}^4 b_i \pd_{\phi_i}$  to vary inside the mesonic torus $\U(1)^4$, and introducing a set of $d$ K\"ahler parameters $\lambda_a$ for $\text{SE}_7$. The latter can be incorporated by deforming $C(\text{SE}_7)$ into a $\U(1)^4$ toric fibration $M_8$ over a polytope $\cP = \{ v^a_i y_i - \lambda_a \geq 0 \}$ in $\bR^4$. The conditions for supersymmetry can be expressed in terms of the equivariant volume \cite{Martelli:2023oqk}
\bea
 \bV (\lambda_a , b_i) = \frac{1}{(2\pi)^4} \int_{M_8} e^{ - b_i y_i} \frac{\omega^4}{4!} \, ,
\eea
where $\omega = \rd y_i \wedge \rd \phi_i$. The equivariant volume depends only on topological data and can be computed by resolving the CY conical singularity, decomposing the fan into a union of tetrahedra $\{ v^{a_1}, v^{a_2}, v^{a_3}, v^{a_4}\}$, and using the localization formula:
\bea
 \bV ( \lambda_a , b_i ) = \sum_{A=\{a_1,a_2,a_3,a_4\}} \frac{e^{-\frac{\phi_A}{d_A}}}{ d_A \prod_{i=1}^4 \epsilon^{(A)}_i} \, ,
\eea
where $d_A = |(v^{a_1}, v^{a_2}, v^{a_3}, v^{a_4})|$, and $\frac{\phi_A}{d_A} = \sum_{i=1}^4 \epsilon^{(A)}_i \lambda_{a_i}$ with $\epsilon^{(A)}_i = \frac{(b, v^{a_j}, v^{a_k}, v^{a_l})}{(v^{a_i}, v^{a_j}, v^{a_k}, v^{a_l})}$ for $i \neq j \neq k \neq l$. Here and in the rest of the paper  $( v^a , v^b , v^c , v^d)$ denotes the determinant of the vectors $v^a$, $v^b$, $v^c$ and $v^d$. The restriction to $\lambda_a = 0$ is the Sasakian volume of $\text{SE}_7$ defined in \cite{Martelli:2005tp,Martelli:2006yb},
\bea
 \bV(0, b_i) = \frac{3}{\pi^4} \Vol_{\text{SE}_7}(b_i) \, ,
\eea
which is a function of the Reeb vector only, and the cubic part in a formal Taylor expansion in $\lambda_a$ is the master volume $\cV (\lambda_a, b_i)$ defined in \cite{Gauntlett:2018dpc}
\bea
 \label{master}
 \cV(\lambda_a,b_i) = ( 2 \pi )^4 \bV(\lambda_a,b_i)\Big |_{\text{cubic terms in}\, \lambda_a} \, .
\eea
The equivariant volume depends on the chosen resolution, but the Sasakian volume does not, and when SE$_7$ is smooth, the master volume is also independent of the resolution. $\cV ( \lambda_a, b_i)$ is a homogeneous function of degree three in $\lambda_a$ and degree minus one in $b_i$. The value of $b_1$ is fixed by supersymmetry, but we will keep it arbitrary for convenience. Only $d - 3$ parameters $\lambda_a$ are independent. Indeed, one can check that \cite{Martelli:2023oqk}
\bea
 \label{shift}
 \bV \Bigl( \lambda_a + \sum_{j=1}^4\beta_j v_j^a,b_i \Bigr) = e^{-\sum_{i=1}^4 \beta_i b_i} \bV(\lambda_a , b_i) \, ,
\eea
for arbitrary $\beta_i \in \bR$. It follows that the $\phi_A$, and therefore $\bV ( \lambda_a, b_i)$ and $\cV (\lambda_a, b_i)$, are invariant under the gauge transformation $\lambda_a \to \lambda_a + \sum_{i=1}^4 \gamma_i (b_1 v_i^a - b_i)$ with $\gamma_i \in \bR$. In particular,
\bea
 \label{conb}
 b_1 \sum_{a=1}^d v_i^a \frac{\partial\cV}{\partial\lambda_a}= b_i \sum_{a=1}^d \frac{\partial\cV}{\partial\lambda_a}\, .
\eea

We are now ready to introduce the generalized free energy. We define it as a \emph{constrained Legendre transform of the master volume}. More precisely, we impose the constraint $N = - \sum_{a=1}^d \frac{\pd \cV}{\pd \lambda_a}$ and define the R-charges as $\Delta_a = - \frac{2}{N} \frac{\pd \cV}{\pd \lambda_a}.$ These satisfy $\sum_{a=1}^d \Delta_a = 2$. The previous relations can be inverted to express the variables $( \lambda_a, b_i)$ in terms of $(\Delta_a, N, b_1)$. In each set, only $d + 1$ parameters are independent. We can then consider the master volume as a function of $\Delta_a$. The function $\cV( \Delta_a)$ is given by $N^{3/2} b_1^{1/2}$ times a homogeneous function of degree two in the variables $\Delta_a$. These relations can be inverted to determine the Reeb vector (the mesonic moduli)
\bea\label{Reeb}
 \frac{2b_i}{b_1}=\sum_{a=1}^d v_i^a \Delta_a \, ,
\eea
which follows from \eqref{conb}, and the baryonic K\"ahler moduli
\bea
 \label{bar}
 X^{(r) }\equiv \sum_{a=1}^d B_a^{(r)} \lambda_a = - \frac{4}{N} \sum_{a = 1}^{d} B_a^{(r)} \frac{\pd \cV ( \Delta_a )}{\pd \Delta_a} \, ,\eea
which has been verified in examples. We also set
\bea
 \label{free}
 F_{S^3}(\Delta_a) \equiv (4 \pi)^3 b_1^{-1/2} \cV(\Delta_a) \, ,
\eea
which should be regarded as a generalization of the large-$N$ free energy, recovering known results when available and offering predictions where they are not. The extremization of $F_{S^3}$ with respect to $\Delta_a$ determines the exact R-charges of the SCFT. Indeed, due to \eqref{bar}, extremizing $F_{S^3}$ with respect to the baryonic directions sets $X^{(r)} = 0$ for $r = 1, \ldots, d - 4$. Purely mesonic variables can be characterized as $\lambda_a = \sum_{i=1}^4 \beta_i v_i^a$, and it then follows from \eqref{shift} that all $\phi_A / d_A = \sum_{j = 1}^4 \beta_j b_j \equiv - \Phi$ are equal, leading to $\cV = ( 2 \Phi)^3 \Vol_{\text{SE}_7}(b_i)$. A short computation using $\Phi \bigl|_{\lambda_a = 1} = - b_1$ then gives $N = 24 b_1 \Phi^2 \Vol_{\text{SE}_7}(b_i)$, and $F_{S^3}(\Delta_a) = \frac{16}{b_1^2} \sqrt{\frac{2\pi^6}{27\Vol_{\text{SE}_7}(b_i)}}$ to be extremized with respect to $b_i$. One can also check that 
\bea
 \label{BZ}
 \Delta_a(b_i) =\frac{2\pi}{3 b_1} \frac{\Vol_{S_a}(b_i)}{\Vol_{\text{SE}_7}(b_i)}\, ,
\eea
where $S_a$ is the torus-invariant five-cycle associated with $v^a$ \footnote{For an alternative derivation see section 4.3 of \cite{Hosseini:2019ddy}.}. All these results are in agreement with \cite{Herzog:2010hf,Jafferis:2011zi,Martelli:2011qj}, where it is also checked that the volume minimization with respect to $b_i$ is equivalent to the field theory $F$-maximization. However, notice that the large-$N$ expression
$F_{S^3}(\Delta_a) \bigl|_{\text{ref. \cite{Jafferis:2011zi}}} = F_{S^3}( \Delta_a) \bigl|_{\text{ours}\, X^{(r)}=0}$
is blind to baryonic directions, and its extremization predicts the exact R-charges of mesonic operators only. In contrast, our $F_{S^3}(\Delta_a)$ explicitly depends on all variables and can be used to predict the R-charges of baryons as well. It is the three-dimensional analogue of the cubic formula \cite{Benvenuti:2006xg} for the large-$N$ central charge, $a (\Delta_a) = \frac{9 N^2}{62} \sum_{a,b,c = 1}^{d} |(v^a, v^b, v^c)| \Delta_a \Delta_b \Delta_c$, for D3-branes probing CY three-folds.  Our formula generalizes and completes the attempts to find a formula based on a quartic expression \cite{Amariti:2011uw,Amariti:2012tj}  by incorporating the baryonic directions. The analogy is complete when we observe that the constrained Legendre transform of the master volume for a five-dimensional Sasaki-Einstein manifold is precisely the large-$N$ trial central charge, $a (\Delta_a)$, as shown in \cite{Hosseini:2019ddy} (see \cite{Chen:2024erz} for a recent application).  

We now consider M-theory black hole horizons with topology AdS$_2 \times \Sigma_\fg \times \text{SE}_7$, which can be realized by further wrapping the set of $N$ M2-branes on a Riemann surface $\Sigma_\fg$ of genus $\fg$ with a topological twist. We allow for integer magnetic fluxes $\fn_a$ along $\Sigma_\fg$ for all global symmetries. The supersymmetric conditions of \cite{Gauntlett:2018dpc} read
\begin{align*}
 & N = - \sum_{a = 1}^d \frac{\pd \cV}{\pd \lambda_a} \, , \\
 & N \fn_a = - \frac{A}{2 \pi} \sum_{ b = 1}^d \frac{\pd^2 \cV}{\pd \lambda_a \pd \lambda_b}
 - b_1 \sum_{i = 1}^{4} n^i \frac{\pd^2 \cV}{\pd \lambda_a \pd b_i} \, ,
\end{align*}
\bea
 \label{susy:constraint}
 & A \sum_{a , b = 1}^d \frac{\pd^2 \cV}{\pd \lambda_a \pd \lambda_b}
 = 2 \pi n^1 \sum_{a = 1}^d \frac{\pd \cV}{\pd \lambda_a}
 - 2 \pi b_1 \sum_{ i = 1}^{4} n^i \sum_{a = 1}^d \frac{\pd^2 \cV}{\pd \lambda_a \pd b_i} \, ,
\eea
where $n^i = \sum_{a=1}^d v_i^a \fn_a$, and $A$ is the area of $\Sigma_\fg$. Supersymmetry also requires $b_1 = 1$ and $\sum_{a = 1}^d \fn_a = 2 - 2 \fg$. We can use the constraints \eqref{susy:constraint} to eliminate $\lambda_a$ and $A$, expressing them as functions of $b_i$ and $\fn_a$. We then define the entropy function as \cite{Gauntlett:2018dpc}
\bea
 \label{entropy:func:main}
 S ( b_i , \fn_a ) \equiv - 8 \pi^2 \bigg( A \sum_{a = 1}^d \frac{\pd \cV}{\pd \lambda_a}
 + 2 \pi b_1 \sum_{i = 1}^{4} n^i \frac{\pd \cV}{\pd b_i} \bigg) \, .
\eea
The entropy function $S ( b_i , \fn_a )$ should be extremized with respect to $b_i$ to find the entropy of the black hole with charges $\fn_a$.

Our main result is the following. By expressing $b_i$ and $\lambda_a$ in terms of $\Delta_a$ and using the constraints \eqref{susy:constraint}, we find
\bea
 \label{susyaction}
 S ( \Delta_a , \fn_a ) = - \frac{1}{2} (4 \pi )^3 \sum_{a = 1}^{d} \fn_a \frac{\pd \cV ( \Delta_a )}{\pd \Delta_a} \, .
\eea
In order to find the entropy of the black hole, we can extremize $S(\Delta_a , \fn_a)$ with respect to $\Delta_a$. It may seem that there are more variables $\Delta_a$ than $b_i$, but it turns out that the generalized entropy function is automatically extremized with respect to the baryonic directions once the constraints \eqref{susy:constraint} are satisfied:
\bea
 \label{barextr}
 \sum_{a = 1}^{d} B_a^{(r)} \frac{\pd S ( \Delta_a , \fn_a )}{\pd \Delta_a}= 0 \, , \qquad r = 1, \ldots , d - 4 \, .
\eea
The function $S ( \Delta_a , \fn_a )$ provides a large-$N$ prediction for the topologically twisted index, depending on all fluxes and fugacities. As shown in \cite{Hosseini:2019ddy,Gauntlett:2019roi}, it matches the known large-$N$ results for the index \cite{Hosseini:2016tor,Hosseini:2016ume} in the special case of a mesonic twist, $X^{(r)} = 0$, which imposes the additional constraint
\bea
 \sum_{a = 1}^{d} B_a^{(r)} \frac{\pd \cV ( \Delta_a )}{\pd \Delta_a} = 0 \, ,
\eea
and leads to R-charges satisfying \eqref{BZ}.

Checking \eqref{bar}, \eqref{susyaction} and \eqref{barextr} is computationally challenging, but we have carried out this analysis in \cite{Hosseini:2025mgf} for the full series of examples considered in \cite{Hosseini:2019ddy}, where we will also provide a general proof. Some of these examples, like many known dualities, involve CY$_4$ whose toric diagram contains external faces with four or more vertices. In such cases, the link is singular, and the master volume depends on the choice of resolution; however, \eqref{bar}, \eqref{susyaction} and \eqref{barextr} hold for all choices. For the mesonic twist, all resolutions yield the same result. It would be interesting to understand the physical meaning of these choices when baryonic fluxes are included. 

\emph{\textbf{An example: the manifold ${\bm Q^{1,1,1}}$.---}}
$Q^{1,1,1}$ is the coset $\SU(2)^3/\U(1)^2$ with toric data:   $v^1=(1, 1 , 0, 0 )$, $v^2= (1, 0, 1, 0 )$, $v^3 = (1, 0, 0, 1 )$, $v^4= (1, 1, 0, 1)$, $v^ 5 = (1, 1 , 1 , 0 )$ and $v^6 = (1, 0, 1, 1 )$.
By removing the common first entry we have the  toric diagram:
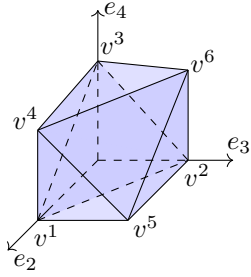
\begin{equation}
\label{toric:Q111}
\begin{tikzpicture}[scale=0.4]

  \draw[draw=none,fill=blue!20,opacity=0.6] (3,0) -- (3,3) -- (0,3.3) -- (-2,1) -- cycle;
  \draw[draw=none,fill=blue!20,opacity=0.6] (3,0) -- (1,-2) -- (-2,1) -- (3,3) -- cycle;
  \draw[draw=none,fill=blue!20,opacity=0.6] (3,0) -- (1,-2) -- (-2,-2) -- (-2,1) -- cycle;
  \draw[draw=none,fill=blue!20,opacity=0.6] (3,0) -- (-2,-2) -- (-2,1) -- (0,3.3) -- cycle;
  \draw[draw=none,fill=blue!20,opacity=0.6] (3,0) -- (3,3) -- (-2,1) -- (-2,-2) -- cycle;

  \draw[-,dashed] (0,0) -- (3,0) node[below] {};
  \draw[->,solid] (3,0) -- (4.5,0) node[below] {};
  
  \draw[-,dashed] (0,0) -- (0,3) node[below] {};
  \draw[->,solid] (0,3) -- (0,5) node[below] {};
  
  \draw[-,dashed] (0,0) -- (-1.6,-1.6) node[below] {};
  \draw[->,solid] (-1.6,-1.6) -- (-3.,-3.) node[below] {};
  
  \draw (-2.4,-3.4) node {$e_2$};
  \draw (4.7,0.5) node {$e_3$};
  \draw (0.6,5) node {$e_4$};
  
  \draw[-,solid] (-2,1) -- (1,-2);
  \draw[-,solid] (-2,1) -- (3,3);
  \draw[-,solid] (1,-2) -- (3,3);
  \draw[-,solid] (-2,-2) -- (1,-2);
  \draw[-,solid] (-2,-2) -- (-2,1);
  \draw[-,dashed] (-2,-2) -- (3,0);
  \draw[-,solid] (3,3) -- (3,0);
  \draw[-,solid] (3,0) -- (1,-2);
  \draw[-,solid] (0,3.3) -- (3,3);
  \draw[-,solid] (0,3.3) -- (-2,1);
  \draw[-,dashed] (-2,-2) -- (0,3.3);
  \draw[-,dashed] (3,0) -- (0,3.3);

  \draw (-1.7,-2.5) node {$v^1$};
  \draw (3.3,-0.5) node {$v^2$};
  \draw (0.5,3.9) node {$v^3$};
  \draw (-2.4,1.4) node {$v^4$};
  \draw (1.6,-2.2) node {$v^5$};
  \draw (3.5,3.3) node {$v^6$};

\end{tikzpicture}
\end{equation}
There are two baryonic symmetries: $B^{(1)} = (1, -1, 0, -1, 0, 1)$ and $B^{(2)} = (0, -1, 1, -1, 1, 0)$. For more details about the geometry, see \cite{Fabbri:1999hw}. The dual SCFT, identified in \cite{Benini:2009qs}, is a flavored ABJM quiver. We already checked the equivalence of $\cI$-extremization and its gravitational dual for the mesonic twist in \cite{Hosseini:2019ddy}, to which we refer for the field theory interpretation of the fugacities. Here, we extend the entropy function to a general twist. We can triangulate the toric polytope \eqref{toric:Q111} by considering the tetrahedra $(2436)$, $(2546)$, $(2514)$, and $(2341)$. The master volume can be computed from \eqref{master} as a sum of four contributions \footnote{This was also computed by different methods in \cite{Hosseini:2019ddy, Cassia:2025aus}.}. Eliminating $\lambda_a$ and $b_i$, it takes the form  
\bea
 \label{V:on-shell:Delta:Q111}
 \cV ( \Delta_a ) & = \frac{N}{3 b_1}
 \frac{\Delta _5 \left( \Delta_1 \phi_3  - \Delta_6 \phi_2  \right)
 - \Delta_3 \left( \Delta_1 \phi_4  - \Delta_6 \phi_1  \right)}
 {( \Delta_3 - \Delta_5 ) ( \Delta_1 - \Delta_6 )} \, ,
\eea
where the $\phi_A$ are determined by
\begin{widetext}
\bea
 \label{phi:Q111}
 \phi_1^2  & = \frac{b_1^3 N}{128 \pi^4}
 ( \Delta_1 + \Delta_2 + \Delta_5 ) ( \Delta_1 + \Delta_4 + \Delta_5 )
 \left( \Delta_6 ( \Delta_3 - \Delta_5 )
 +  \frac{128 \pi^4\phi_2^2}{b_1^3 N( \Delta_1 + \Delta_2+ \Delta_3 ) ( \Delta_1 + \Delta_3 + \Delta_4 )} \right) , \\
 \phi_3^2 & = \frac{b_1^3 N}{128 \pi^4}
 ( \Delta_2 + \Delta_3 + \Delta_6 ) ( \Delta_3 + \Delta_4 + \Delta_6 )
 \left( \Delta_5 ( \Delta_1 - \Delta_6 )
 +  \frac{128 \pi^4\phi_2^2 }{b_1^3 N( \Delta_1 + \Delta_2 + \Delta_3 ) ( \Delta_1 + \Delta_3 + \Delta_4 )} \right) , \\ 
 \phi_4^2 & = \frac{b_1^3 N}{128 \pi^4}
 ( \Delta_2 + \Delta_5 + \Delta_6 ) ( \Delta_4 + \Delta_5 + \Delta_6 )
 \left( \Delta_1 \Delta_3 - \Delta_5 \Delta_6
 + \frac{128 \pi^4\phi_2^2}{b_1^3 N( \Delta_1 + \Delta_2 + \Delta_3 ) ( \Delta_1 + \Delta_3 + \Delta_4 )} \right) ,
\eea
\end{widetext} 
and the constraint
$ \phi_1  - \phi_2  + \phi_3  - \phi_4  = 0$.
The baryonic K\"ahler moduli are given by $X^{(1)}  = - \frac{2}{b_1}  \frac{\phi_2  - \phi_3 }{\Delta_1 - \Delta_6}$ and $X^{(2)} =- \frac{2}{b_1} \frac{\phi_3  - \phi_4}{\Delta_3 - \Delta_5}$. By introducing the quartic function as in \cite{Amariti:2011uw,Amariti:2012tj}
\bea
 \label{a3d:Xr:def:Q111}
 a_{(3)} ( \Delta_a ) & \equiv \! \sum_{a , b , c , e= 1}^{6} c_{abcd}\Delta_a \Delta_b \Delta_c \Delta_e
 - ( \Delta_2 \Delta_4 )^2 - ( \Delta_3 \Delta_5)^2 \\
 & - ( \Delta_1 \Delta_6 )^2
 + \frac{( \Delta_2 \Delta_4 + \Delta_3 \Delta_5 + \Delta_1 \Delta_6 )^2}{2} \, ,
\eea
where $c_{abcd} = \frac1{24}| ( v^a , v^b , v^c , v^e ) |$, \eqref{V:on-shell:Delta:Q111} can be rewritten as
$
 \cV ( \Delta_a ) = \pm \frac{N^{3/2} \sqrt{b_1}}{24 \sqrt{2} \pi^2}
 \sqrt{ a_{(3)}  + Y } \, ,
$
with
\bea
 \frac{b_1 N}{2 (2 \pi )^4} Y & \equiv ( X^{(1)} - X^{(2)} ) \left( \Delta_1 \Delta_6 X^{(1)} - \Delta_3 \Delta_5 X^{(2)} \right) \\
 & + \Delta_2 \Delta_4 X^{(1)} X^{(2)} \, .
\eea
For the mesonic twist, $X^{(1)} = X^{(2)} = 0$, and thus $ \cV_{\text{mes.}} ( \Delta_a ) = \pm \frac{N^{3/2} \sqrt{b_1}}{24 \sqrt{2} \pi^2} \sqrt{a_{(3)}}$, which is consistent with \cite[(5.47)]{Hosseini:2019ddy}. 

\emph{\textbf{A purely baryonic twist.---}}
Using the $\SU(2)^3$ symmetry of the model, it is also possible to consistently truncate to a purely baryonic twist by setting $\Delta_1 = \Delta_6 \equiv \delta_1$, $\Delta_2 = \Delta_4 \equiv \delta_2$, and $\Delta_3 = \Delta_5 \equiv \delta_3$, along with $\fn_1 = \fn_6 \equiv \fp_1$, $\fn_2 = \fn_4 \equiv \fp_2$, and $\fn_3 = \fn_5 \equiv \fp_3$. The R-charge constraint reduces to $\sum_{a = 1}^{3} \delta_a = 1$, and the twisting condition to $\sum_{a = 1}^{3} \fp_a = 1 - \fg$. Using \eqref{Reeb}, we find that the mesonic variables are frozen to their canonical values: $b_2 = b_3 = b_4 = \frac{1}{2} b_1$. By taking a suitable limit in the previous formulae, we find $\phi_l = \pm \frac{b_1^{3/2} \sqrt{N}}{(4 \pi )^2} ( \delta_1+ \delta_2 + \delta_3)^{3/2} \sqrt{\frac{ \delta_1 \delta_3}{ \delta_2}}$ for $l = 1 , \ldots, 4$ and  $\cV_{B} ( \delta_a ) = \mp \frac{\sqrt{b_1} N^{3/2}}{(4 \pi)^2} \sqrt{\delta_1 \delta_2 \delta_3 ( \delta_1 + \delta_2 + \delta_3 )}$. It is extremized at $\delta_a = \frac{1}{3}$, which correctly leads to the known dimension of baryons realized as M5-branes wrapped on the five-cycles associated with $v^a$ \cite{Fabbri:1999hw,Hanany:2008fj}. The entropy function is then
$ S_B ( \delta_a , \fp_a )  = - \frac{1}{2} ( 4 \pi )^3 \sum_{a = 1}^{3} \fp_a \frac{\pd \cV_B ( \delta_a )}{\pd \delta_a}$ and can be used to predict the entropy of purely magnetic black holes. For dyonic black holes with electric charges $\fq_a$, we would extremize
\bea
 \label{dyonic}
 \cI_B ( \delta_a ) \equiv S_B ( \delta_a , \fp_a ) + 2 \pi \ii \sum_{a = 1}^{3} \delta_a \fq_a \, .
\eea

This result can be compared to the entropy of asymptotically AdS$_4 \times Q^{1,1,1}$ dyonic BPS black holes found in \cite{Halmagyi:2013sla}. These solutions are described within the framework of four-dimensional gauged $\cN = 2$ supergravity, coupled to $n_{\text{V}} = 3$ vector multiplets and $n_{\text{H}} = 1$ hypermultiplet, which is a consistent truncation of M-theory on $Q^{1,1,1}$ \cite{Cassani:2012pj}. The truncation includes, in addition to the graviphoton dual to the R-symmetry, two massless vector fields arising from the reduction of the M-theory four-form on three-cycles, which are thus dual to baryonic symmetries. The fourth vector is massive. The dynamics of the theory are fully specified by the prepotential, the Killing vector, and the Killing prepotential. We use the notations of \cite{Halmagyi:2013sla} to which we refer for further details. The prepotential governing the vector multiplet sector is given by $F ( X^{\Lambda} ) = 2 \sqrt{X^0 X^1 X^2 X^3}$ where $X^{\Lambda}$, $\Lambda = 0, \ldots, 3$, are the projective coordinates on the scalar manifold. The gauging of the theory is purely electric and is determined by the Killing vector $k_{\Lambda}^a = \sqrt{2} ( e_0 , 2, 2, 2 )$ acting on the hypermultiplet scalar fields $(a , \phi, \zeta , \tilde \zeta)$, and the corresponding Killing prepotential $P_{\Lambda}^{3} = \sqrt{2} \bigl( 4 - \frac{e_0}{2} e^{2\phi} , - e^{2\phi} , -e^{2\phi} , - e^{2\phi} \bigr)$, $P^{3\, ,\Lambda} = 0$, where $e_0$ is related to the AdS$_4$ radius by $R_{\text{AdS}_{4}} = \frac{1}{2} \left( \frac{e_0}{6} \right)^{3/4}$. We consider dyonic black holes with magnetic and electric charges $p^\Lambda$ and $q_\Lambda$, respectively. Supersymmetry requires $p^\Lambda P_\Lambda^3 = \mp 1$ and $p^\Lambda k_\Lambda^a = 0$. The BPS condition can be expressed as attractor equations \cite{DallAgata:2010ejj}, which are equivalent to extremizing
\bea
 \label{attractor}
 \cI ( X^\Lambda) \equiv - \ii \frac{\Vol_{\Sigma_\fg}}{4  G_{\text{N}}} \frac{  q_\Lambda X^\Lambda - p^\Lambda F_\Lambda }{ P^3_\Lambda X^\Lambda - P^{3, \Lambda} F_\Lambda} \, ,
\eea
with respect to the projective coordinates $X^\Lambda$, where $F_\Lambda \equiv \frac{\pd_\Lambda F(X^\Lambda)}{\pd X^\Lambda}$. In this notation, extremizing \eqref{attractor} with respect to $X^\Lambda$ yields the entropy of the dyonic black hole with charges $p^\Lambda$ and $q_\Lambda$. The attractor equation \eqref{attractor} needs to be supplemented by the hyperino BPS variation evaluated at the horizon, $X^\Lambda k^a_\Lambda = 0$, which implies the constraint $e_0 X^0 + 2 (X^1 + X^2 + X^3) = 0$. This condition can be used to eliminate $X^0$ (and the corresponding massive vector field) and to define an \emph{effective prepotential} for the massless vector fields, $F_{\text{eff.}}(X^\Lambda) = 2 \sqrt{\frac{2}{e_0}} \sqrt{- X^1 X^2 X^3 (X^1 + X^2 + X^3)}$, which can then be inserted into \eqref{attractor}. Varying the original equation \eqref{attractor} with respect to $X^0$ fixes the horizon values of the dilaton $\phi$ and the non-conserved electric charge $q_0$. Notice that eliminating the field $A^0$ leads to a redefinition of the corresponding electric charges: $\wh q_a = q_a - \frac{2}{e_0} q_0$, $a=1,2,3$. The attractor equations in this form have been used extensively to compare gravity and field theory in various examples, including \cite{Benini:2016rke,Benini:2017oxt,Hosseini:2017fjo,Hosseini:2018uzp,Hosseini:2018usu,Bobev:2018uxk,Benini:2020gjh,Hosseini:2020wag}. By restricting the extremization problem to the fields $X^1, X^2, X^3$, identifying the quantized charges
\bea
 \frac{2 \sqrt{2}\,\Vol_{\Sigma_{\fg}}}{\pi e_0}\, p^a \in \bZ \, , \qquad \frac{e_0 \Vol_{\Sigma_\fg}}{64 \sqrt{2} \pi G_{N}} \, \wh q_a \in \bZ \, ,
\eea
with  $\fp_a$ and $\fq_a$, respectively, and setting: $\frac{X^a}{\sum_{b = 1}^{3} X^b} = \delta_a$, we precisely recover the field theory extremization \eqref{dyonic}.
We used the holographic relation $\frac{R_{\text{AdS}_{4}}^2}{G_{\text{N}}} = \frac{2 \sqrt{6} \pi^2}{9} \frac{N^{3/2}}{\sqrt{\Vol_{Q^{1,1,1}}}}$
with $\Vol_{Q^{1,1,1}} = \frac{\pi^4}{8}$.

\emph{\textbf{Reduction to ${\bm M^{1,1,1}}$.---}}
We note that, at the level of the truncation discussed in \cite{Cassani:2012pj}, results for the manifold $M^{1,1,1}$ can be obtained by further truncating the $Q^{1,1,1}$ model through the identifications $X^3 = X^1$ and $p^3 = p^1$, $q^3 = q^1$. For $M^{1,1,1}$ (as for any chiral quiver), there are no large-$N$ field theory results available; however, we can still employ our formulae to compute the entropy function. For brevity, we simply quote the final result here. The large isometry group $\SU(3) \times \SU(2)$ of $M^{1,1,1}$ allows us to set the Reeb vector to a canonical value and consistently consider a purely baryonic twist. We find $\cV_{B}(\delta_a) = \mp \frac{\sqrt{b_1}\,N^{3/2}}{\sqrt{2}(4\pi)^2}\,\delta_1\,\sqrt{\delta_2(3\delta_1 + 2\delta_2)}$, subject to the constraint $3 \delta_1 + 2 \delta_2 = 2$. It is extremized at $\delta_1=\frac{4}{9}$ and $\delta_2=\frac{1}{3}$, which correctly leads to the known dimension of baryons realized as M5-branes wrapped on the five-cycles associated with $v^a$ \cite{Fabbri:1999hw,Hanany:2008fj}. We explicitly see that $\cV_{B} (\delta_a)$ can be obtained by restricting and appropriately redefining the fugacities in the purely baryonic master volume computed for $Q^{1,1,1}$, as anticipated by gravity. One can then easily verify that this result precisely reproduces the entropy of the asymptotically AdS$_4 \times M^{1,1,1}$ dyonic BPS black holes found in \cite{Halmagyi:2013sla}. The same result was previously obtained in \cite{Kim:2019umc} by explicitly analyzing the supersymmetry conditions of \cite{Gauntlett:2018dpc}.

\emph{\textbf{Discussion and Outlook.---}}
The main result of this letter is the proposal for a universal gravitational block that controls the Bekenstein-Hawking entropy of all AdS$_4$ BPS black holes in M-theory, in the philosophy of \cite{Hosseini:2019iad}. The block is defined as a constrained Legendre transform, $\cV (\Delta_a)$, of the master volume of the internal geometry. In this letter, we have analyzed only magnetically charged static black holes, and it would be interesting to introduce rotation or to find classes of Kerr-Neumann or accelerating black holes that are computationally tractable. We expect that all extremization problems for asymptotically AdS$_4 \times \text{SE}_7$ black holes can be rewritten in terms of $\cV (\Delta_a)$. On the field theory side, $\cV (\Delta_a)$ should also appear in the asymptotics of the giant graviton expansion for the superconformal index of M2-brane theories, as conjectured in \cite{Chen:2024erz}. More broadly, it provides an explicit prediction for the large-$N$ limit of the $S^3$ partition function for cases where existing methods fail to compute it.

\begin{acknowledgements}
SMH is supported by UK Research and Innovation (UKRI) under the UK government's Horizon Europe funding guarantee [grant number EP/Y027604/1]. AZ is partially supported by the INFN, and the MIUR-PRIN grant No. 2022NY2MXY [finanziato dall'Unione europea -- Next Generation EU, Missione 4 Componente 1 CUP H53D23001080006].
\end{acknowledgements}

\bibliography{BarBH}

\end{document}